\documentclass[12pt, twoside, a4paper]{article}
\usepackage[latin1]{inputenc}
\usepackage[T1]{fontenc}
\usepackage{supertabular}
\usepackage{graphics}
\usepackage{epsfig}
\usepackage{bbm}
\usepackage{latexsym}
\usepackage[intlimits]{amsmath}
\usepackage{amssymb}
\usepackage{psfrag}
\usepackage{rotating}
\usepackage{mathrsfs}
\usepackage{setspace}
\usepackage[dvips]{color}

\usepackage{geometry}
\newcommand{\dau}[1]{\partial_{#1}}

\newcommand{\Dau}{\mathcal{D}}
\newcommand{\de}{\mathrm{d}}

\newcommand{\commute}[2]{\left[#1,#2\right]}
\newcommand{\rez}[1]{\frac{1}{#1}}
\newcommand{\tr}[1]{\mathrm{tr}\left(#1\right)} 
\newcommand{\btr}[1]{\mathrm{Tr}\left(#1\right)}   
\newcommand{\trp}[1]{\mathrm{Tr}'\left(#1\right)}  
\newcommand{\varexp}[1]{\mathrm{exp}\left(#1\right)}

\newcommand{\bra}[1]{\langle #1|}
\newcommand{\ket}[1]{|#1\rangle}

\newcommand{\zvec}[1]{{\mathbf #1}}
\newcommand{\ha}{\hspace{1cm}}

\newcommand{\img}{\mathrm{im}}

\renewcommand{\Im}{\mathrm{Im~}}

\newcommand{\id}{{\mathbbm 1}}

\newcommand{\Om}{\Omega}
\newcommand{\Et}{U}
\newcommand{\G}{G}
\newcommand{\F}{\mathscr{F}}
\newcommand{\T}{T}

\newcommand{\Lat}{\mathscr{L}}
\newcommand{\rt}{\zvec{m}}
\newcommand{\zt}{\zvec{k}}
\newcommand{\pfn}{\mathrm{Pf}(n_{\mu\nu})}
\newcommand{\Z}{\mathscr{Z}}
\newcommand{\D}{\hat D}
\newcommand{\Zs}{\mathscr{Z}}
\newcommand{\M}{M}
\newcommand{\dA}{\delta A}
\newcommand{\Q}{Q}
\newcommand{\q}{q}
\newcommand{\ce}{c}
\newcommand{\be}{b}

\newcommand{\len}{L}
\newcommand{\laq}{\Lambda_{QCD}}
\newcommand{\so}{\Theta}

\newcommand{\ab}{\tilde a}
\newcommand{\bb}{\tilde b}

\newcommand{\pol}{\mathscr{P}}

\newcommand{\RR}{\mathbbm{R}}
\newcommand{\TT}{\mathbbm{T}}
\newcommand{\ZZ}{\mathbbm{Z}}

\newcommand{\Ao}{\mathrm{A}^{(0)}}
\newcommand{\Go}{\mathrm{G}^{(0)}}

\title{Free Energy of thick Center Vortices\footnote{supported 
       by DFG under grant-No. DFG-Re
       856/4-2}}

\author{Ch.~Korn, H.~Reinhardt and T.~Tok\\
        Institut f\"ur Theoretische Physik\\
	Universit\"at T\"ubingen\\
	Auf der Morgenstelle 14}

\begin{document}

\maketitle
\onehalfspacing

\abstract{The free energy of thick center vortices is calculated 
in continuum Yang-Mills theory in 
one-loop approximation using the proper time regularization. The 
vortices are represented by Abelian gauge field configurations on 
the torus which satisfy twisted boundary conditions.}

\section{Introduction}

Understanding color confinement 
in the framework of QCD is one of the most challenging 
non-perturbative problems of strong interaction. 
Center vortices offer an appealing picture of confinement 
\cite{'tHooft:1978hy,Mack:1979rq,Greensite:2003bk}. 
The vortex picture of the QCD vacuum was proposed as early as in 1978
and has recently received support by lattice calculations.
After center projection in the so called maximal center gauge one
produces still the full string tension \cite{DelDebbio:1997mh}. 
On the other hand, when center vortices are removed from the 
Yang-Mills ensemble the string tension vanishes \cite{deForcrand:1999ms}. 
In addition, the signals of confinement disappear from QCD propagators in
both the Landau gauge \cite{Gattnar:2004bf} and the Coulomb gauge
\cite{Greensite:2004ke} when the center vortices are removed.
Furthermore, the deconfinement phase transition in the vortex picture 
can be understood as a depercolation transition \cite{Engelhardt:1999fd}
and the topological charge can be understood in terms of intersections
\cite{Engelhardt:1999xw} and writhing \cite{Reinhardt:2001kf} of the 
center vortex sheets. 

In refs.~\cite{kovacs:00,deForcrand:2001nd} the free energy of center vortices 
has been investigated in lattice calculations exploiting the fact that
on the torus a center vortex can be induced by twisted boundary
conditions. In ref.~\cite{kovacs:00} it was shown that 
in the confinement regime the free energy of a (thick) center vortex 
vanishes. In ref.~\cite{deForcrand:2001nd} the t'~Hooft loop 
\cite{'tHooft:1977hy}, which generates a center vortex 
\cite{Reinhardt:2002mb}, has been used to calculate the free energy 
in the sectors of $SU(2)$ Yang-Mills theory with fixed electric flux 
as a function of temperature and spatial volume. 
In the present paper we will carry out an analogous
investigation in continuum Yang-Mills theory. We will introduce the
center vortices on the torus by means of twisted boundary conditions.
These boundary conditions are then realized by Abelian background
fields. We calculate the free energy of such center vortex fields as
function of the temperature in one-loop approximation\footnote{In 
this context let us also mention that the energy density of an
infinitely thin static magnetic center vortex in continuum $SU(2)$ 
Yang-Mills theory has been evaluated in \cite{Lange:2003ti} in the 
Schr\"odinger picture to one loop order.}.

The paper is organized as follows: In the next section we summarize 
the relevant features of gauge fields on the torus satisfying  
twisted boundary conditions. In particular we focus on the 
derivation of the free energy of configurations with definite 
electric flux and its relation to the free energy of a static 
quark-antiquark pair. In section \ref{vortex_field} we introduce 
Abelian configurations fulfilling twisted boundary conditions. 
Such configurations can be interpreted as electric or magnetic 
center vortices. Then the operator of fluctuations around these gauge 
fields is determined. 
The spectrum of the fluctuations around such 
background gauge fields with {\em non-zero} instanton number has been 
calculated already in ref.~\cite{vbaal:84} and in general it has 
negative modes. However, positivity of the fluctuation operator 
can be achieved by restriction to {\em zero} instanton number and by a 
suitable choice of some free parameters (moduli) of the background 
gauge fields. This fact, which was pointed out in \cite{vbaal:96} for gauge 
fields on the three torus, can be trivially generalized to the four 
torus for gauge fields with zero Pontryagin index. Then the requirement 
of a strictly positive fluctuation operator constrains 
the range of temperatures. For the cases
of purely spatial and purely temporal twists (where the Pontryagin 
index vanishes) the spectrum of the fluctuation operator is calculated. 
Finally, in section \ref{calc_free_energy}, the proper time
regularization is used to calculate the free energy of magnetic and
electric vortices to one-loop order. Our results are then confronted 
with those of the lattice calculations of ref.~\cite{deForcrand:2001nd}.

\section{QCD on the hypertorus\label{torusqcd}}

To fix our notation we summarize the relevant ingredients of gauge fields 
on the torus $\TT^4$. We define the four torus $\TT^4$ as $\RR^4$ 
modulo the lattice
\begin{align}
        \label{eq:02_0042}
 \Lat = \left\{x \in {\RR}^4 | x = n_\mu \len^{(\mu)}; n =
 (n_\mu), n \in 
     {\mathbbm Z}^4 \right\} \, ,
\end{align}
where $\len^{(\mu)} := \len_\mu e_\mu\, , \, \mu = 0,1,2,3$ 
denote the vectors spanning the lattice and 
$\{e_\mu| (e_\mu)_\nu = \delta_{\mu\nu}\}$ is the
canonical basis of $\RR^4$. Throughout this paper we choose
$L_1=L_2=L_3=L$ and $L_0=1/T$ with $T$ being the temperature. 
Fields on $\TT^4$ are considered then as fields on $\RR^4$ fulfilling 
appropriate boundary conditions.
Local gauge invariants are periodic with respect to a shift by an
arbitrary lattice vector $\len^{(\mu)}$, whereas the gauge potential 
is (in general) only periodic up to a gauge transformation, i.e.
\begin{align}
        \label{eq:02_0045}
        A_\lambda(x + \len^{(\mu)}) =A_\lambda^{\Om_\mu}(x)
        = \Om_\mu(x)A_\lambda(x)\Om_\mu(x)^{\dagger} +
        \Om_\mu(x)\dau{\lambda}\Om_\mu(x)^{\dagger}.
\end{align}
Here $\Om_\mu (x)$ denotes the transition function in $\mu$-direction.
Transition functions in different directions have to respect the cocycle
condition
\begin{align}
        \label{eq:02_0047}
        \Om_\mu(x + \len^{(\nu)})\Om_\nu(x) =
           Z_{\mu\nu}\Om_\nu(x + \len^{(\mu)})\Om_\mu(x) \, ,
\end{align}
where 
\begin{align}
        \label{eq:02_0048}
        Z_{\mu\nu} = \varexp{\frac{2\pi i}{N} n_{\mu\nu}}\cdot\id_N,
        \ha n_{\mu\nu} \in {\mathbb Z}(\mathrm{mod}~ N)
\end{align}
is an element of the center of the gauge group $SU(N)$. Here we have 
introduced the antisymmetric \emph{twist-tensors} $n_{\mu\nu}$ and
$\id_N$ is the $N \times N$ unit matrix. The cocycle condition (\ref{eq:02_0047}) 
expresses compatibility of two successive translations in the 
$(\mu,\nu)$-plane, and a non-trivial twist $Z_{\mu\nu}$ induces 
a center vortex in this plane. The possible twists are divided into 
two groups. First, the spatial twists $n_{i j} , (i,j = 1,2,3)$ which 
can be interpreted as the components of a 3-vector 
$\rt = (n_{23},n_{31},n_{12})$ which represents the direction of the
magnetic flux. These twists induce magnetic center vortices. Second, 
temporal twists $n_{0 i}$, which again can be interpreted as the
components of a 3-vector $\zt = (n_{01},n_{02},n_{03})$. These 
twists induce ``fat'' electric center vortices whose flux is
homogeneously distributed over the whole torus. \\
\noindent
The Pontryagin index $P_N$ of a gauge potential is given by 
\begin{align}
        \label{eq:02_0049}
        P_N = - \rez{16\pi^2}\int_{\TT^4}\de^4 x \,\,\tr{\G_{\mu\nu}
        \G_{\mu\nu}^*} \, , \, 
\end{align}
where 
\begin{align}
        \label{eq:02_0008}      
        \G_{\mu\nu} = \dau\mu A_\nu - \dau\nu A_\mu +
        \commute{A_\mu}{A_\nu} = \G^a_{\mu\nu}\T_a 
\end{align} 
is the field strength, $\G_{\mu\nu}^* =
\rez{2}\epsilon_{\mu\nu\alpha\beta}\G_{\alpha\beta}$ its dual and $\T_a$
are the generators of the Lie algebra of the gauge group normalized such
that $\tr{\T_a \T_b} = -\frac{1}{2} \delta_{a b}$. The
Pontryagin index is fully determined by the transition functions
$\Om_\mu$. In \cite{vbaal:82} it has been shown that the Pontryagin
index is generally a fractional number
\begin{align}
        \label{eq:02_0049a}
        P_N = \nu + \left(\frac{N-1}{N}\right)\pfn 
	\, , \, \nu  \in \ZZ \, , 
\end{align}
where $\nu$ is the integer valued topological charge (instanton number) 
and $\pfn = \rez{8} \epsilon_{\mu\nu\alpha\beta}n_{\mu\nu}n_{\alpha\beta}$
is the Pfaffian of the twist tensor $n_{\mu\nu}$. The fractional 
part of $P_N$ is obviously due to $n_{\mu \nu}$. 
Under a gauge transformation $\Et$ the transition functions $\Om_\mu$
transform as
\begin{align}
   \label{eq:02_0052}
   \Om^{\Et}_\mu(x) = \Et(x + \len^{(\mu)})\Om_\mu(x)\Et^{-1}(x).
\end{align}
The twist tensor $n_{\mu\nu}$ is invariant under gauge transformations
of the transition functions.

\subsection{Flux}
\label{flux}

In this subsection we will derive the free energy of a 
pure $SU(N)$ gauge field configuration on $\TT^4$ with given 
electric and magnetic flux $(\zvec{e}, \rt)$. Thereby we will 
mainly follow the ideas outlined by 't~Hooft \cite{thooft:79}. 

The partition function under interest is given by the trace over physical
states with well defined electric and magnetic flux $(\zvec{e}, \rt)$. 
This partition function can be formally expressed as
\begin{align}
        \label{eq:02_0064}
        \Z(\zvec{e}, \rt,\len_\mu) :=
        \mathcal{N} \btr{{\mathsf
        P}(\zvec{e},\rt)e^{-\beta H}} \, ,
\end{align}
where ${\mathsf P}(\zvec{e},\rt)$ denotes the projector on definite
electric and magnetic flux, $\beta = L_0$ is the inverse temperature 
and $\mathcal{N}$ is a 
normalization constant chosen such that $\Z(\zvec{e}=0, \rt,\len_\mu)=1$.
To obtain this projector we will first identify the physical
states with given flux $(\zvec{e}, \rt)$. To this end we consider the
canonically quantized theory (in Weyl gauge $A_0 = 0$). 
In this gauge the configuration space consists of spatial gauge potentials 
$A_i \, , \, i= 1,2,3$ on the 
torus $\TT^3$ satisfying the twisted boundary conditions
(\ref{eq:02_0045}) with spatial twist vector $\rt$. This vector $\rt$
represents the magnetic flux of the configuration under consideration. 
Since electric field
operators do not commute with the gauge potential, it is harder
to identify states with fixed electric flux $\zvec{e}$. 
Physical states have to be invariant under (time-independent) gauge
transformations $\Om(\zvec{x})$. The gauge group of pure $SU(N)$
Yang-Mills theory is $SU(N)/Z(N)$, since the gauge potentials
live in the adjoint representation. Therefore, gauge 
transformations take values in $SU(N)/Z(N)$ rather than in $SU(N)$, 
i.e.~a gauge transformation $\Om(\zvec{x})$ need not be periodic on 
$\TT^3$ - but can change across the torus by a center element:
\begin{align}
        \label{eq:02_0053}
        \Om[\zt](\zvec{x + \len^{(i)}}) = Z_i\Om[\zt](\zvec{x}),\ha i =
        1,2,3,\ha Z_i := e^{\frac{2\pi i k_i}{N}}\id_N \in Z(N) \, . 
\end{align} 
Correspondingly, since $\pi_1(SU(N)/Z(N)) = Z(N)$, on $\TT^3$ 
there are $N^3$ homotopically inequivalent classes of gauge transformations. 
The different classes are labeled by a vector $\zvec{k} =
(k_1,k_2,k_3) \in {\mathbbm{Z}}^3(\!\!\!\mod N)$ and transformations 
belonging to a class $\zvec{k}$ are denoted by $\Om[\zt](\zvec{x})$. 
One can 
choose a representative $\tilde \Om[\zt](\zvec{x})$ from each class 
which takes values only in the Cartan subgroup of $SU(N)$ such that
\begin{align}
        \label{eq:02_0054A}
	\tilde \Om[\zt_1] \tilde \Om[\zt_2] = 
	\tilde \Om[(\zt_1+\zt_2)\!\!\!\mod N] \, .
\end{align}
A general gauge transformation of class $\zt$ is then given by 
the product of $\tilde \Om[\zt](\zvec{x})$ and a topologically 
trivial (i.e. truly periodic) gauge transformation (belonging to 
class $\zt=0$). Physical states $\ket{\psi}$ 
have to be gauge invariant under topologically trivial (or small) gauge
transformations, but they can pick up a phase under a topologically
non-trivial gauge transformation. 
Let $\hat\Om[\zt](\zvec{x})$ be the unitary operator which, when 
acting on the Hilbert space of physical states, generates the gauge 
transformation $\tilde \Om[\zt](\zvec{x})$. Obviously, the operators 
$\hat\Om[\zt](\zvec{x})$ with different $\zt$ commute with 
each other and with the Hamiltonian. The eigenstates of 
$\hat\Om[\zt](\zvec{x})$ possess well defined electric flux 
as can be seen as follows. Since the $\hat\Om[\zt](\zvec{x})$ 
are unitary their eigenvalues are pure phase
\begin{align}
        \label{eq:02_0054}
        \hat\Om[\zt]\ket{\psi} = e^{i\sigma(\zt)}\ket{\psi},\ha
        \sigma(\zt)\in{\mathbbm R} \, .
\end{align}
and from eq.~(\ref{eq:02_0054A}) follows
that the phase factor $\sigma$ is of the form
\begin{align}
        \label{eq:02_0055}
        \sigma(\zt) = \frac{2\pi}{N}\zvec{e}\cdot\zt
\end{align}
with some vector $\zvec{e}\in{\mathbbm{Z}}^3(\!\!\!\!\!\mod N)$,
i.e. we can label the eigenstate $\ket{\psi}$ by the vector 
$\zvec{e}$, i.e.~$\ket{\psi_{\zvec{e}}}$. We will now identify 
the vector $\zvec{e}$ as the
electric flux. For this purpose we notice that the Wilson loop operator 
\begin{align}
        \label{eq:02_0056}
        \hat W(C) = 
	\rez{N}\tr{ \mathcal{P} \varexp{-\oint_{C}A_\mu\de x^\mu}} 
\end{align}
is the creation operator of electric flux lines \cite{Polyakov:1975rs}. 
Let $C_i,~ i = 1,2,3$ denote a path linearly interpolating between 
points $\zvec{x}$ and $\zvec{x} + \len^{(i)}$.
The corresponding Wilson loop operator $\hat W(C_i)$ is not 
invariant under homotopically non-trivial gauge transformations 
$\Om[\zt]$:
\begin{align}
        \label{eq:02_0058}
        \hat W(C_i)^{\Om[\zt]} = \rez{N}\tr{\Om[\zt](\zvec{x}+
        \len^{(i)})\mathcal{P}
        e^{-\int_{C_i}\zvec{A}\de\zvec{x}}\Om[\zt]^\dagger(\zvec{x})}
        = Z_i \hat W(C_i)  \, ,
\end{align}
\noindent
where $Z_i$ is the center element defined in (\ref{eq:02_0053}). 
Consider now the action of $\hat\Om[\zt]$ on the state
$\hat W(C_i)\ket{\psi_{\zvec{e}}}$:
\begin{align}
        \label{eq:02_0059}
	\hat \Om[\zt] \hat W(C_i)\ket{\psi_{\zvec{e}}}
	=
	\hat W(C_i)^{\Om[\zt]}
	\ket{\psi^{\Om[\zt]}_{\zvec{e}}}
	=
	Z_i \hat W(C_i)\hat\Om[\zt] \ket{\psi_{\zvec{e}}} 
	= 
	e^{i\frac{2\pi}{N}(\zvec{e}\cdot \zt + k_i)}
	\hat W(C_i)\ket{\psi_{\zvec{e}}} \, .
\end{align}
It is seen that the action of $\hat W(C_i)$ on $\ket{\psi_{\zvec{e}}}$ 
increases the $i$-th component of the vector ${\zvec{e}}$ by one unit. 
Since the Wilson loop operator is the creation operator of electric 
flux, we have to identify $\frac{2 \pi}{N} {\zvec{e}}$ with the 
electric flux vector. Accordingly the eigenvectors 
$\ket{\psi_{\zvec{e}}}$ (\ref{eq:02_0054}) of $\hat \Om[\zt]$ are 
states with definite electric flux $\zvec{e}$ and the projection 
operator on physical states with given electric flux ${\zvec{e}}$ 
is given by 
\begin{align}
        \label{eq:02_0064B}
        {\mathsf P}(\zvec{e}) = 
	\rez{N^3}\sum_{\zt}
	e^{-\frac{2\pi i}{N}\zvec{e}\zt} \hat\Om[\zt] \,  .
\end{align}
Indeed acting with ${\mathsf P(\zvec{e})}$ on the state 
$\ket{\psi_{\zvec{e}'}}$ one obtains:
\begin{align}
        \label{eq:02_0064C}
        {\mathsf P}(\zvec{e}) \ket{\psi_{\zvec{e}'}}= 
	\rez{N^3}\sum_{\zt}
	e^{-\frac{2\pi i}{N}\zvec{e}\zt} \hat\Om[\zt] 
	\ket{\psi_{\zvec{e}'}} = 
	\rez{N^3}\sum_{\zt}
	e^{-\frac{2\pi i}{N}\zvec{e} \zt} 
	e^{\frac{2\pi i}{N}\zvec{e}' \zt} 	
	\ket{\psi_{\zvec{e}'}} = 
	\delta^{(3)}_{\zvec{e} , \zvec{e}'} 
	\ket{\psi_{\zvec{e}'}} \, .
\end{align}
The desired projector ${\mathsf P}(\zvec{e},\rt)$ on definite 
electric and magnetic flux $(\zvec{e},\rt)$ is then given by 
${\mathsf P}(\zvec{e})$ times the projector onto states with fixed 
spatial twist $\rt$. With the projector ${\mathsf P}(\zvec{e},\rt)$ 
at our disposal the partition function for fixed electric and 
magnetic flux $(\zvec{e},\rt)$ defined by eq.~(\ref{eq:02_0064}) 
is given by
\begin{align}
        \label{eq:02_0065a}
	\Z(\zvec{e},\rt) = 
	\frac{\mathcal{N}}{N^3}\sum_{\zt}
	e^{-\frac{2\pi i}{N} (\zvec{e}\cdot\zt)} 
	\Z_{\zt} (\rt) = 
	\frac{\sum_{\zt} e^{-\frac{2\pi i}{N} (\zvec{e}\cdot\zt)} 
	\Z_{\zt}(\rt)}{\sum_{\zt} \Z_{\zt}(\rt)} \, ,
\end{align}
\noindent
where
\begin{align}
        \label{eq:02_0065b}
	\Z_{\zt}(\rt) = 
	\btr{\hat\Om[\zt]e^{-\beta H}}
\end{align}
is the partition function of fixed temporal twist $\zt$. (Recall that
fixed spatial twist $\rt$ implies also fixed magnetic flux.)
Eq.~(\ref{eq:02_0065a}) defines the $Z(N)$-Fourier transform from the
temporal twist $\zt$ to the electric flux $\zvec{e}$ and is refered to
as Kramers-Wannier duality.

Formally the partition function of fixed temporal twist $\zt$ 
(\ref{eq:02_0065b}) represents the thermodynamic average of the operator
$\hat\Om[\zt]$ generating gauge transformations which are periodic in
spatial direction up to a center element, see eq.~(\ref{eq:02_0053}). 
In the ``coordinate'' representation this partition function is given
by (ignoring for the moment the spatial twist)
\begin{align}
        \label{eq:02_0065c}
	\Z_{\zt} &=
	\int \Dau \zvec{A} \bra{\zvec{A}} 
	\hat\Om[\zt]e^{-\beta H} P_G \ket{\zvec{A}}
	\nonumber \\
	&=
	\int \Dau \zvec{A} \bra{\zvec{A}^{\Om[\zt]}} 
	e^{-\beta H} P_G \ket{\zvec{A}} \, ,
\end{align}
where $P_G$ denotes the projector on gauge invariant states 
\cite{Reinhardt:1997rm}. The matrix element in the above equation can be
expressed in the standard form by a functional integral over spatial
gauge field configurations satisfying the temporal boundary condition
\begin{align}
        \label{eq:02_0065d}
	\zvec{A} (\vec{x},\beta) = 
	\zvec{A}^{\Om[\zt]} (\vec{x},0) \, .
\end{align}
Furthermore the projector $P_G$ contains an integration over the gauge
group with the ``Haar'' measure. This integral can be expressed as an
integral over a (temporally constant) temporal gauge field $A_0$. Thereby
the Haar measure becomes the Faddeev-Popov determinant in the gauge
$\partial_0 A_0 = 0$. The resulting functional integral for the
partition function (\ref{eq:02_0065c}) is gauge invariant and can be
expressed in an arbitrary gauge yielding
\begin{align}
        \label{eq:02_0065e}
	\Z_{\zt} = 
	\int_{\zvec{A} (\beta) = \zvec{A}^{\Om} (0)} 
	\Dau A_\mu \delta_{gf} (A) e^{-S_{YM}(A)} \, ,
\end{align}
where the spatial gauge field satisfies the temporally twisted boundary
conditions (\ref{eq:02_0065d}). Adding also spatially twisted boundary
conditions introduces spatial flux $\rt$ as described at the beginning 
of the section. This shows that the partition function 
(\ref{eq:02_0065b}) can be calculated from the standard functional 
integral supplemented by the twisted boundary conditions 
(\ref{eq:02_0053}).

\subsection{Polyakov-loop}

The static $q\overline{q}$-potential at finite temperature 
$T = \rez{\beta}$ is defined as the alteration of the free energy 
$F_{q\overline{q}}(\beta)$ after adding a $q\overline{q}$-pair 
to the vacuum \cite{Svetitsky:1985ye}. 
This potential can be extracted from the Polyakov-loop correlator
\begin{align}
        \label{eq:02_0068}
        \langle \pol(\zvec{x})\pol^\dagger(\zvec{y})\rangle
	=
        e^{-\beta F_{q\overline{q}}(\zvec{x},\zvec{y})} \, ,
\end{align}
\noindent   
where in presence of temporal twist $\Om_0(x)$ the gauge invariant 
Polyakov loop operator is defined by
\begin{align}\label{eq:02_0067}
        \pol(\zvec{x}) 
	=
	\rez{N} \tr{{\mathcal P}\varexp{-\int_0^{\beta=\len_0} 
        {A_0}(x^0,\zvec{x})\de x^0} \Om_0^\dagger(\zvec{x})} 
	\, . 
\end{align}
\noindent
Using the boundary conditions (\ref{eq:02_0045}) for the gauge potential
and the cocycle condition (\ref{eq:02_0047}) one obtains 
the following periodicity properties of the Polyakov loop:
\begin{align}
        \pol(\zvec{x} + \len^{(i)}) &=\nonumber
	\rez{N}\tr{\mathcal{P}
	e^{-\int_0^\beta \de x^0 A_0(x^0, \zvec{x} + \len^{(i)})} 
	\Om_0^\dagger(\zvec{x} + \len^{(i)})}\\
        \label{eq:02_0069}
	&=
        \rez{N}\tr{\mathcal{P}
	e^{-\int_0^\beta \de x^0 \Om_i(A_0 + \dau{0})\Om_i^\dagger}
	Z_{i0}\Om_i(x)\Om_0^\dagger(\zvec{x})
        \Om_i^\dagger(x+\beta)}=e^{-\frac{2\pi 
        i}{N}k_i}\pol(\zvec{x}) \, ,
\end{align}
\noindent 
where $k_i$ is the $i$-th component of the temporal twist vector $\zt$.
After multiple usage of this periodicity property one arrives at 
\begin{align}
        \label{eq:02_0070}
        \pol(\zvec{x} - \len\zvec{e}) = e^{\frac{2\pi
        i}{N}\zt\cdot\zvec{e}}\pol(\zvec{x})
\end{align}
\noindent
From this relation follows that the thermic Polyakov loop correlator 
in presence of magnetic flux $\zvec{m}$ is given by
\begin{align}
        \label{eq:02_0071}
        \langle\pol(\zvec{x})
        \pol^\dagger(\zvec{x}-\len\zvec{e})\rangle = 
        \frac{\sum_{\zt}e^{-\frac{2\pi i}{N}(\zvec{e}\cdot\zt)}
	      \Z_{\zt}(\rt)}{\sum_{\zt}\Z_{\zt}(\rt)}
\end{align}
\noindent
Comparison with eq.~(\ref{eq:02_0065a}) shows that
\cite{deForcrand:2001nd}
\begin{align}
        \label{eq:02_0071a}
        \langle\pol(\zvec{x})
        \pol^\dagger(\zvec{x}-\len\zvec{e})\rangle = 
	\Z(\zvec{e},\rt) \equiv
	e^{-\beta F(\zvec{e},\rt)} \, .  
\end{align}
\noindent
Hence, on the torus the implementation of temporal twisted boundary 
conditions enforces static electric flux which can be interpreted as
arising from two homogeneously but oppositely charged planes a distance
$L {\zvec{e}}$ apart, and thus simulates static quark and antiquark
sources a distance $L {\zvec{e}}$ apart (Note that $\langle\pol(\zvec{x})
\pol^\dagger(\zvec{x}-\len\zvec{e})\rangle$ is independent of
$\zvec{x}$). In the following we will realize the twisted boundary
conditions by Abelian background fields.

\section{Abelian vortex field}
\label{vortex_field}

In the remaining part of the paper we restrict ourselves to 
the gauge group $SU(2)$. 
The implementation of flux on the torus by means of twisted boundary
conditions can be most easily realized by Abelian background gauge
fields\footnote{Here, Abelian gauge potential means a potential that 
takes values only in the Cartan subalgebra of the 
gauge group.}  of constant field strength on $\TT^4$. Such 
a configuration can be interpreted as the field of a fat  
center vortex whose flux is homogeneously smeared out over the 
whole torus. 

The most general solution of the equations of motion of pure $SU(2)$
Yang-Mills theory with constant field
strength and twisted boundary conditions on $\TT^4$ reads 
\cite{vbaal:84} 
\begin{align}\label{eq:03_0008}
        {\Ao_\mu}(x) =
        \left(\frac{\pi}{2}\F_{\mu\nu}x_\nu - 
	\frac{\pi \Q_\mu}{\len_\mu}\right)\T,\ha \T :=
        -i\left(\begin{array}{cc}1&0\\0&-1
	                 \end{array}\right) 
	= -i\sigma_3 \, ,
\end{align}
where $\Q_\mu \, , \, \mu  = 0,1,2,3$ are arbitrary constants (moduli) 
and 
\begin{align*}
\F_{\mu\nu} = - \frac{n_{\mu \nu}}{\len_\mu \len_\nu}
\end{align*}
is up to constant factor the field strength 
\begin{align}\label{eq:03_0009}
        \Go_{\mu\nu} = -\pi \F_{\mu\nu} \T 
\end{align}
induced by the twist tensor $n_{\mu \nu}$. 
The gauge field (\ref{eq:03_0008}) fulfills the twisted boundary 
conditions (\ref{eq:02_0045},\ref{eq:02_0047}) with transition functions
\begin{align}
        \label{eq:03_0015}
        \Om_\mu (x) =
        \varexp{\frac{\pi}{2} \sum_\nu n_{\mu\nu} \frac{x_\nu}{\len_{\nu}}\T} 
\end{align}
and twist tensor $n_{\mu\nu}$.  
Obviously, each non-zero component $n_{\mu \nu} = - n_{\nu \mu}$ 
of the twist tensor corresponds to a non-zero field strength component
$\Go_{\mu\nu}$ of the gauge potential (\ref{eq:03_0008}), 
which represents the field of $|n_{\mu \nu}|$ center vortices 
in the $(\mu \nu)$-plane homogeneously smeared out on $\TT^4$. 
We will use this field to calculate the free 
energy of ``fat'' center vortices in one-loop order. To this end we 
will compute first the spectrum and then the determinant of the 
operator of fluctuations 
of the gauge field around the given center vortex configuration 
(\ref{eq:03_0008}). The appearance of negative eigenvalues will be
avoided by appropriately choosing the constants $\Q_\mu$.

\subsection{Fluctuation operator}
\label{fluctuation_operator}

The spectrum of the fluctuations around the Abelian potential 
(\ref{eq:03_0008}) for non-zero Pontryagin index (\ref{eq:02_0049a}) 
has been found already in \cite{vbaal:84} and in general it has 
negative modes. On the three torus these negative modes can be avoided 
\cite{vbaal:96} by appropriately choosing the moduli. Here we will
extend this result to the four torus where one has to restrict oneself
to background gauge fields with zero Pontryagin index and to 
choose the constants $\Q_\mu$ (moduli) (\ref{eq:03_0008}) appropriately. 
We will shortly summarize the essential results needed below.

The fluctuation operator is obtained by expanding the Yang-Mills action 
around the background field $\Ao_\mu$ (\ref{eq:03_0008}).  
For this purpose it is convenient to express the fluctuating gauge
potential as
\begin{align}
        \label{eq:03_0022}
        A_\mu = \Ao_\mu + \dA_\mu \, , \ha 
	\dA_\mu = \frac{-i}{2}
        \left(
                \begin{array}{cc}
                        \be_\mu & \sqrt{2} \ce_\mu \\
                        \sqrt{2} \ce_\mu^* &  - \be_\mu
                \end{array}
        \right) \, .
\end{align}
\noindent
From eqs.~(\ref{eq:02_0045},\ref{eq:03_0015}) 
the boundary conditions for $\be_\mu$ and
$\ce_\mu$ follow:
\begin{align}
    \label{eq:03_0024}
    \be_\mu(x + \len^{(\lambda)}) = \be_\mu(x) \, , \ha 
    \ce_\mu(x + \len^{(\lambda)}) = 
      \varexp{-\pi i \sum\limits_\nu \frac{n_{\lambda\nu}x_\nu}{\len_\nu}}
      \ce_\mu(x) \, . 
\end{align}
\noindent 
For later use we write down the transformation properties of 
$\ce_\mu$ under a shift by an arbitrary lattice vector 
$ l_\nu \len^{(\nu)} $ which can be obtained by successive 
use of eq.~(\ref{eq:03_0024}):
\begin{align}
        \label{eq:03_0026}
       \ce_\mu\left(x + \sum_\nu l_\nu \len^{(\nu)}\right) =
        \varexp{-i\pi\sum_{\rho,\nu}\len_{\rho}l_\rho 
        \F_{\rho\nu}x_\nu + i\pi
        \sum_{\rho<\nu}l_\rho n_{\rho\nu}l_\nu}\ce_\mu(x).
\end{align}
\noindent
Noting that $\Ao_\mu$ is a solution of the equations of motion the
action in terms of the fluctuations $\dA_\mu$ reads
\begin{align}
        \label{eq:03_0027}
        S_{YM} = S_0 & -\rez{g^2} \int \de^4 x \left \{
        \tr{-2\dA_\mu\commute{\Go_{\mu\nu}}{\dA_\nu} 
        - \dA_\mu(\D_\nu^0)^2 \dA_\mu 
         -(\D_\mu^0 \dA_\mu)^2 + \mathcal{O}(\dA^3)}\right\} \, ,
\end{align}
\noindent 
where $S_0 = -\rez{2g^2} \int \de^4 x~ \tr{{\Go_{\mu\nu}}^2}$ is the
action of the background field $\Ao_\mu$ and 
\begin{align}
  \label{eq:03_0006}
  \D_\mu^0 := \dau\mu +   \commute{\Ao_\mu}{\cdot}
\end{align}
is the covariant derivative with respect to $\Ao_\mu$ in the
adjoint representation. Adopting background gauge fixing
\begin{align}
  \label{eq:03_0007}
  \D_\mu^0 \dA_\mu = 
  \dau\mu \dA_\mu + \commute{\Ao_\mu}{\dA_\mu}
  \stackrel{!}{=} 0
\end{align}
\noindent
one finds for the gauge fixed action 
\begin{align}
        \label{eq:03_0028}
        S^{gf} &= S_0 -
        \rez{g^2}\!\int\!\!\de^4x
	\left\{\tr{\dA_\mu\M_A^{\mu\nu}\dA_\nu} +
        2\tr{\bar\Psi\M_{gh}\Psi} + 
	\mathcal{O}(\dA^3)\right\} \, 
\end{align}
with the fluctuation operators
\begin{align}
        \label{eq:03_0029}
        \M_A^{\mu\nu} = 
	- \delta_{\mu\nu}(\D_\lambda^0)^2 - 
	2  \commute{\Go_{\mu\nu}}{\cdot} \, , \ha 
        \M_{gh} = -(\D_\lambda^0)^2 \, 	
\end{align}
and $\Psi$ being the ghost field. Using the parametrization
(\ref{eq:03_0022}) and representing the ghost field as
\begin{align}
        \label{eq:03_0030}
        \Psi = \frac{-i}{2}\left(\begin{array}{cc}\eta&2\sqrt{2}\phi\\
        2\sqrt{2}(\phi')^*&-\eta\end{array}\right)
\end{align}  
\noindent 
one finally obtains up to terms of third order in the fluctuations 
\begin{align}
         \label{eq:03_0031}
        S^{gf} &
         &= S_0 + \rez{g^2}\int\de^4x
	 \left\{\rez{2}\be_\mu\M_0\be_\mu +
        \ce_\mu^*\left(\M_n\delta_{\mu\nu} - 
	4\pi i \F_{\mu\nu} \right)\ce_\nu + 
	\eta^*\M_0\eta + 
	\phi^*\M_n\phi +
        {\phi'}^*M_n\phi' \right\} \, ,
\end{align}
\noindent
where 
\begin{align}
        \label{eq:03_0032}
        \M_0 := \left(\rez{i}\dau \lambda\right)^2\!\!\! , \ha
        \M_n := \left(\rez{i}\dau\lambda - \pi\F_{\lambda\nu}x_\nu +
        \frac{2\pi\Q_\lambda}{\len_\lambda}
	\right)^2 \, .
\end{align}
\noindent
Obviously, the operator $\M_n$ depends on the twist tensor 
$n_{\mu \nu} = - \frac{\F_{\mu\nu}}{\len_\mu \len_\nu}$.
Note that the ghost fields $\phi , \phi'$ and $\eta$ respect 
the same periodicity properties as the gauge fields $\ce_\mu$ 
and $\be_\mu$, respectively, see eq.~(\ref{eq:03_0024}).

\subsection{The spectrum of the fluctuation operator}
\label{spectrum}

In this section we will calculate the spectrum of the operators $\M_0 ,
\M_n$ and $\left(\M_n\delta_{\mu\nu}-4\pi i \F_{\mu\nu} \right)$,
see eq.~(\ref{eq:03_0032}), from which the eigenvalues of the 
operators $\M_A$ and $\M_{gh}$ (\ref{eq:03_0029}) follow by using 
the decompositions (\ref{eq:03_0022}) and (\ref{eq:03_0030}).

The eigenfunctions of $\M_0$ are plane waves and  
from the periodicity properties (\ref{eq:03_0024}) of the fields 
$\be_\mu$ and $\eta$ one obtains the eigenvalues  
\begin{align}
\label{eq:03_0033}
\lambda_{l} = \sum_{\mu = 0}^3\left(\frac{2\pi
l_\mu}{\len_\mu}\right)^2,\;\;\;l \in {\mathbbm Z}^4 \, .
\end{align}
\noindent
These eigenvalues are also eigenvalues of $\M_A$ (\ref{eq:03_0029})
(four-fold degenerate in the index $\mu$) 
and of $\M_{gh}$ (non-degenerate).

In the general case the operator $\M_A$ has negative eigenvalues
implying that the considered background gauge potential $\Ao_\mu$ 
is a saddle point of the action. Note that the negative eigenmodes occur
already for covariantly constant background fields, where they signal
the instability of the perturbative vacuum. Only if the map 
\begin{align}
        \label{eq:03_0017}
        \F: \RR^4 \longrightarrow \RR^4 \, , \ha 
        x_\mu \mapsto \F_{\mu\nu}x_\nu \ha  
\end{align}
is degenerate (i.e. $\ker(\F)$ is non-trivial) and for suitable choice 
of the parameters $\Q_\mu$, see eq.~(\ref{eq:03_0008}),  
the spectrum of $\M_n$ is strictly positive. Therefore, to ensure 
positivity, we will consider only the cases of purely spatial (magnetic) 
($\rt \neq 0 , \zt = 0$) and 
purely temporal ($\rt = 0 , \zt \neq 0$) twists. In these cases the map 
$\F$ is obviously degenerate and we have the orthogonal
decomposition\footnote{This decomposition is orthogonal with respect to
the canonical scalar product in $\RR^4$ because the $4\times 4$ matrix
$\F$ is anti-symmetric.}
\begin{align*}
\RR^4 = \ker(\F) \oplus \img(\F) \, ,
\end{align*}
and the operator $\M_n$ decays into two parts 
\begin{align*}
        \M_n & =
	\left.\M_{n} \right|_{\img(\F)} + 
	\left.\M_{n} \right|_{\ker(\F)} \, ,
\end{align*}
\noindent  
each acting in one of the orthogonal spaces $\ker(\F)$ and $\img(\F)$. 
In $\left.\M_{n} \right|_{\ker(\F)}$ the linear term 
$\pi\F_{\lambda\nu}x_\nu$ is absent (because $(x_\mu) \in \ker(\F)$). 
Therefore, the eigenfunctions of $\left.\M_{n} \right|_{\ker(\F)}$ are
plane waves and can be labeled by two integers $k,l$. We denote the
eigenvalues by $\lambda_{(k ,l)}$ and the eigenfunctions by $\ket{k, l}$. 
Eigenfunctions of $\left.\M_{n} \right|_{\img(\F)}$ 
can be labeled by one integer $m$ and are
denoted by $\ket{m}$ with eigenvalue $\lambda_{m}$. Then the 
eigenfunctions of $\M_n$ can simply be written as products of 
eigenfunctions of $\left.\M_{n} \right|_{\img(\F)}$ and 
$\left.\M_{n} \right|_{\ker(\F)}$:
\begin{align*}
        \ket{m,k,l} &= \ket{m}\ket{k, l}\;\;\; \Rightarrow\;\;\; \\
        \M_{n}\ket{m,k,l} &= 
	\ket{k,l} \left.\M_{n} \right|_{\img(\F)} \ket{m} +
        \ket{m} \left.\M_{n} \right|_{\ker(\F)} \ket{k,l} = 
	\left( \lambda_{m} + \lambda_{(k,l)} \right) \ket{m,k,l} \, .
\end{align*}
\noindent 
The spectra of the operators $\left.\M_{n} \right|_{\img(\F)}$ and
$\left.\M_{n} \right|_{\ker(\F)}$ are calculated in Appendix 
\ref{app_spec} from which the spectra of the operators $\M_A$ and
$M_{gh}$ follow immediately and are summarized in tables \ref{tab02} 
and \ref{tab03}.
\begin{table}[h]
    \begin{center}
        \begin{tabular}{l|c|r}
          eigenvalue&degeneracy&parameters\\
          \hline
          $-2\pi f + \left( \zvec{p}(k,l) 
	  - \zvec{\q} \right)^2 $ 
	  & $2\tilde e$ & 
	  $k,l \in {\mathbbm Z}\vphantom{\Big)}$
	  \\ 
          $\hphantom{-}2\pi f + \left( \zvec{p}(k,l)
	  - \zvec{\q} \right)^2 $ & 
	  $6\tilde e$ & 
	  $k,l \in {\mathbbm Z}\vphantom{\Big)}$
	  \\ 
          $2\pi f(2n + 1) + \left(\zvec{p}(k,l)
	  - \zvec{\q} \right)^2 $ & 
	  $8\tilde e$ & 
	  $n = 1,2,3,\ldots; k,l \in {\mathbbm Z}\vphantom{\Big)}$
	  \\ 
          \hline
          from eq.~(\ref{eq:03_0033}):
          $\sum\limits_{\mu=0}^3\left(\frac{2\pi l_\mu}{\len_\mu}\right)^2$
	  & $4$ &
          $l \in {\mathbbm Z}^4\vphantom{\Big)}$\\ 
          \hline
        \end{tabular}
   \caption{\label{tab02} Spectrum of the operator $\M_A^{\mu\nu}$,
   where $\zvec{\q}:= - \left . \frac{2\pi \Q_\lambda}{\len_\lambda}
   \right|_{\ker(\F)}$. The quantities $f$, $\tilde e$ and $\zvec{p}(k,l)$ 
   are defined respectively in eqs.~(\ref{eq:001}), (\ref{eq:002}) and 
   (\ref{eq:03_0054}) of appendix \ref{app_spec}.}
   \end{center}
\end{table}
\begin{table}[h]
    \begin{center}
        \begin{tabular}{l|c|r}
          eigenvalue&degeneracy&parameters\\
          \hline
          $2\pi f(2n + 1) + \left(\zvec{p}(k,l)
	  - \zvec{\q} \right)^2 $ 
	  & $2\tilde e$ & 
	  $n = 0,1,2,\ldots; k,l \in {\mathbbm Z}\vphantom{\Big)}$ 
	  \\ 
          \hline
          from eq.~(\ref{eq:03_0033}):
          $\sum\limits_{\mu=0}^3\left(\frac{2\pi l_\mu}{\len_\mu}\right)^2$
	  & $1$ &
          $l \in {\mathbbm Z}^4\vphantom{\Big)}$\\ 
          \hline
        \end{tabular}
   \caption{\label{tab03} Spectrum of the operator $\M_{gh}$
            (see also the caption of table \ref{tab02}).}
   \end{center}
\end{table}
\noindent

\subsection{No twist}

For sake of completeness let us consider the case without flux 
$\rt=\zt=0$. We will need this spectrum to normalize the free energy of a
twist configuration with respect to the case of zero field strength. 

The eigenvalues of $\M_0$ are the same as in the twisted case, i.e. 
\begin{align}
\label{eq:03_0059}
\lambda_{l} = \sum_{\mu = 0}^3 \left(\frac{2\pi l_\mu
}{\len_\mu} \right)^2,\;\;\;l \in {\mathbbm Z}^4 \, , 
\ha l = (l_\mu) 
\end{align}
\noindent
and they appear in the spectrum of $\M_A^{\mu \nu}$ (with degeneracy 4) 
and of $\M_{gh}$ (non-degenerate).

The remaining spectrum consists of the eigenvalues of the operator 
$\M_n = \left(-i\dau\lambda + 
\frac{2\pi\Q_\lambda}{\len_\lambda}\right)^2$ which read
\begin{align}
\label{eq:03_0060}
\lambda_{l} = \sum_{\mu = 0}^3 
\left(\frac{2\pi }{\len_\mu} (l_\mu-Q_\mu)\right)^2
\, , \;\;\;l \in {\mathbbm Z}^4 \, , 
\ha l = (l_\mu) \, .
\end{align}
These eigenvalues appear in the spectrum of the operator $\M_A^{\mu \nu}$
(with degeneracy 8) as well as of the operator $\M_{gh}$ 
(with degeneracy 2).

\section{Calculation of the free energy}
\label{calc_free_energy}

In this section we will evaluate the free energy of a ``fat'' center
vortex induced by the twisted boundary conditions and 
examine its dependence on the torus geometry,
and especially on the temperature $T=1/L_0$. 
For this purpose we have to evaluate the determinants of the fluctuation
operators. These determinants are ultraviolet singular and need
regularization. We shall adopt the proper-time regularization which
preserves gauge invariance. In the proper time regularization the
determinant of an operator $A$ is given by
\begin{align}
        \label{eq:05_0002}
        \log\det A := -
        \int \limits^\infty_{\rez{\Lambda^2}} \frac{\de \tau}{\tau}
        \trp{\varexp{-\tau A}} = - \int \limits^\infty_{\rez{\Lambda^2}}
        \frac{\de \tau}{\tau} \sum_{\lambda_i\neq 0}\varexp{-\tau\lambda_i}
	\, ,
\end{align}
\noindent
where $\Lambda$ is the ultraviolet cut-off. The symbol $\trp{}$ means 
that the trace is taken over non-zero eigenvalues only.
The $\Lambda^4$ and $\Lambda^2$ divergencies cancel in the ratio of
determinants - in our case the ratio between the determinants for 
non-zero ($n_{\mu\nu}\neq 0$) and zero  ($n_{\mu\nu}= 0$) field
strength. 

The proper time integral (\ref{eq:05_0002}) is defined only if all 
eigenvalues are positive. This requires (see the first line in 
table \ref{tab02}) 
\begin{align}
        \label{eq:05_0035}
        2\pi f < \left(\zvec{p}(k,l) -
        \zvec{\q}\right)^2,\ha \forall k,l \in {\mathbbm Z}. 
\end{align}
\noindent
With given twist tensor and parameters $\Q_\mu$ from this inequality 
follows a condition for $\len T$, which can be interpreted as the
temperature in units of $1/\len$. Therefore, large values $\len T$ 
stand for high temperatures. 

After a lengthy calculation, which is performed in appendix 
\ref{proper_time}, one obtains the free energy of a ``fat'' center
vortex in terms of the renormalized coupling constant:
\begin{align}
        \label{free-en}
	\frac{F_{\zt}(\rt , \len T)}{T} &= 
	\frac{2\pi^2}{g_R^2(T^2)} 
	\frac{\left(f \len^2\right)^2}{\len T}
        + \frac{11}{12} \frac{\left(f \len^2\right)^2}{\len T} 
	\log\frac{1}{L^2 T^2} 
	+ C (f \len^2, \Q_\mu, \len T) \, ,
\end{align}
where the function $C$ is defined in eq.~(\ref{eq:11_0004}) 
of appendix \ref{proper_time}.
In the following we will separately discuss the cases 
of purely spatial (magnetic) twist $\rt$ and purely temporal twist $\zt$. 

\subsection{Spatial twist}
\label{spatial_twist}

\begin{figure}[!t]
  \textsf{
  \begin{center}
        \psfragscanon
        \psfrag{m = (1,0,0)}{$\rt = (1,0,0)$}
        \psfrag{m = (1,1,0)}{$\rt = (1,1,0)$}
        \psfrag{m = (1,1,1)}{$\rt = (1,1,1)$}
        \psfrag{0}{}
        \psfrag{3.5}{}
        \psfrag{4.5}{}
        \psfrag{0.5}{}
        \psfrag{1.5}{}
        \psfrag{2.5}{}
        \psfrag{20}{}
        \psfrag{40}{}
        \psfrag{60}{}
        \psfrag{80}{}
       \psfrag{\len T (Temperatur)}{\Large{$\len T$ (temperature)}}
        \psfrag{Crr(\len T, ||m||)}{\Large{$C(\len T, \|\rt\|)$}}
        \includegraphics[angle = 270,scale=.5]{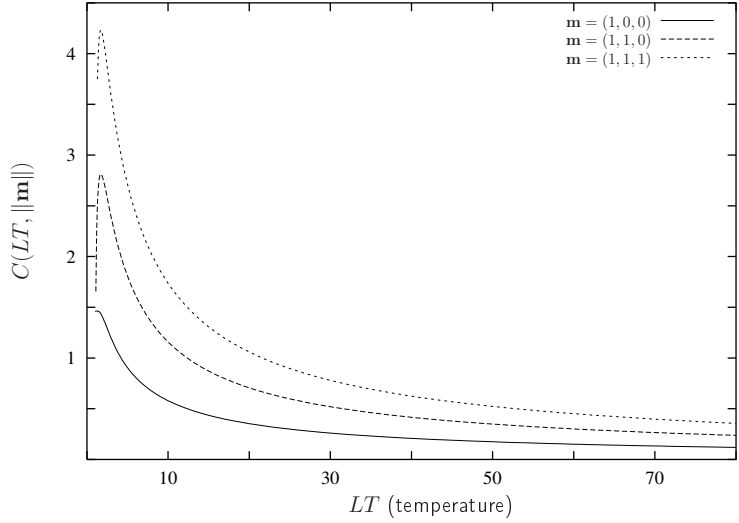}
        \caption{$C$ (\ref{eq:11_0004}) as a function of $\len T$ 
	for different twist vectors.\label{abbtwists}}
  \end{center}
  }
\end{figure}

\begin{figure}[!t]
  \textsf{
  \begin{center}
        \psfragscanon
        \psfrag{m = (1,0,0)}{$\rt = (1,0,0)$}
        \psfrag{m = (1,1,0)}{$\rt = (1,1,0)$}
        \psfrag{m = (1,1,1)}{$\rt = (1,1,1)$}
        \psfrag{0.5}{}
        \psfrag{1.5}{}
        \psfrag{5}{}
        \psfrag{15}{}
        \psfrag{25}{}
        \psfrag{35}{}
        \psfrag{temperature}{\Large{$\frac{T}{\laq}$ (temperature)}}
        \psfrag{freeenergy and more}{\Large{$\frac{F(\rt,
        T/\laq)}{\laq}$ (free energy)}} 
        \includegraphics[angle = 270,scale=.5]{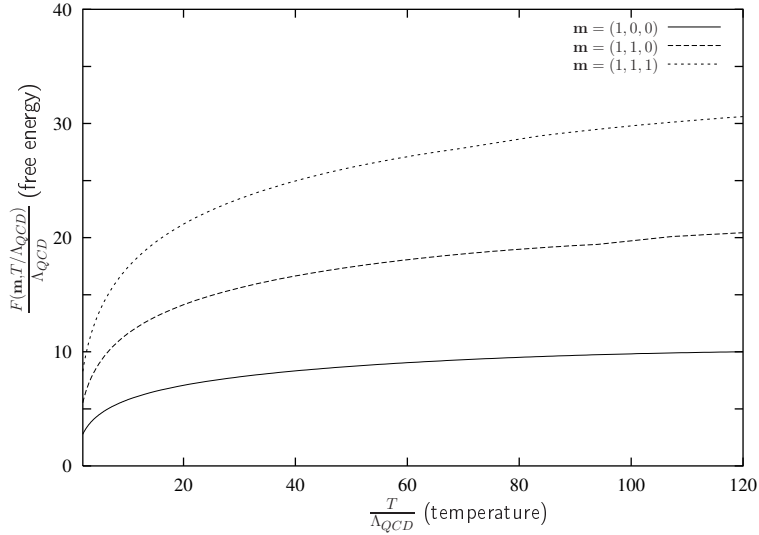}
        \caption{The free energy of thick vortices as a function of
	temperature. \label{abbtwists2}}
  \end{center}
  }     
\end{figure}

For spatial twists $\zt = 0\, , \, \rt \neq 0$, which
correspond to fat magnetic center vortices, we choose 
$\Q_\mu = (\rez 2,\rez 2,\rez 2,\rez 2)$.
In the following table we list the values of the parameter 
$\zvec{\q}= - \left.\frac{2\pi \Q_\lambda}{
\len_\lambda}\right|_{\ker(\F)}$ 
needed in the calculation of the function 
$C$ (\ref{eq:11_0004}) as well as 
the limits for $\len T$ resulting from inequality (\ref{eq:05_0035}):
\begin{equation}
	\begin{array}{c|c|c}
	\rt & \zvec{\q} & 
	\len T
	\\
        \hline
	(1,0,0) & 
	\frac{\pi}{\len}(\len T,1,0,0) & 
	\len T \in {\RR}_+
	\\ 
	(1,1,0) & 
	\frac{\pi}{\len}(\len T,1,1,0) &
	\len T > 0.95
	\\        
	(1,1,1) & 
	\frac{\pi}{\len}(\len T,1,1,1) &
	\len T > 1.05
        \end{array}
\end{equation}
In figure \ref{abbtwists} the quantity $C$ (\ref{eq:11_0004}) is 
plotted as a function of $\len T$ for the different values of $\rt$. 
One observes that $C$ is nearly proportional to the
number of twists (more precisely: proportional to $\|\rt\|^2$).
In figure \ref{abbtwists2} the free energy is plotted in 
units of $\laq$ as a function of $T/\laq$ for the different twist 
configurations, where the torus extension $\len$ has been chosen to be 
$\len = 1/\laq$. The 1-loop expansion is valid if 
$T \gg \laq = \frac{1}{\len}$, i.e.~$\len T \gg 1$. 

\noindent
For large $\len T$ the free energy of thick magnetic center vortices 
is nearly independent of the temperature $T$. Hence, these vortices 
cannot be relevant for the deconfinement phase transition.
In fact it is well known that magnetic center vortices (generated here
by spatial twists) do not contribute to the confining properties of the
theory (i.e.~to the temporal string tension). They do, however,
contribute to the spatial string tension, which even slightly increases
across the deconfinement phase transition. Lattice calculations
\cite{Engelhardt:1999fd} show that magnetic vortices (measured in a 
spatial volume at
fixed time) do percolate in both the confined and deconfined phase and
thus cannot be used to characterize the deconfinement phase transition.

\subsection{Temporal twists}
\label{temporal_twist} 

For electric center vortices induced by temporal
twists $\zt \neq 0 \, , \, \rt = 0$ we 
choose $\Q_\mu = (0,0,\rez 2,\rez 2)$\footnote{Here we have chosen 
$\Q_\mu$ differently from the purely spatial twist case since for 
$\zt=(1,1,1)$ 
the vector $\Q_\mu=(\rez 2,\rez 2,\rez 2,\rez 2)$ would be in the image 
of the map $\F$, see eq.~(\ref{eq:03_0017}), and therefore in
this case we would have $\zvec{\q}=0$.}. 
In the following table 
the parameter $\zvec{\q}$ necessary for the 
calculation of the free energy is
listed. We also quote the range of temperature $\len T$ (last column)
for which all eigenmodes of the fluctuation operator are positive, see
(\ref{eq:05_0035}):
\begin{equation}
	\begin{array}{c|c|c}
	\zt & 
	\zvec{\q} & 
	\len T
	\\
        \hline
	(1,0,0) & 
	\frac{\pi}{\len}(0,0,1,1) &
	\len T < 3.14
	\\
	(1,1,0) & 
	\frac{\pi}{2 \len}(0,-1, 1, 2) &
	\len T < 1.67
	\\
	(1,1,1) & 
	\frac{\pi}{3\len}(0,-2,1,1) &
	\len T < 0.60
	\end{array} \qquad .
\end{equation}
\noindent 
\begin{figure}[!t]
  \textsf{      
  \begin{center}
        \psfragscanon
        \psfrag{k = (1,0,0)}{$\zt = (1,0,0)$}
        \psfrag{k = (1,1,0)}{$\zt = (1,1,0)$}
        \psfrag{k = (1,1,1)}{$\zt = (1,1,1)$}
        \psfrag{0}{$0$}
        \psfrag{1}{$1$}
        \psfrag{2}{$2$}
        \psfrag{-4}{$-4$}
        \psfrag{-3}{$-3$}
        \psfrag{-2}{$-2$}
        \psfrag{-1}{$-1$}
        \psfrag{0.5}{$0.5$}
        \psfrag{1.5}{$1.5$}
        \psfrag{2.5}{$2.5$}
        \psfrag{3.5}{$3.5$}
        \psfrag{3}{$3$}
        \psfrag{4}{$4$}
        \psfrag{6}{$6$}
        \psfrag{8}{$8$}
        \psfrag{10}{$10$}
        \psfrag{LT (Temperatur)}{\Large{$\len T$ (temperature)}}
        \psfrag{Crz(LT, ||k||)}{\Large{$C(\len T, \|\zt\|)$}}
        \includegraphics[angle = 270,scale=.66]{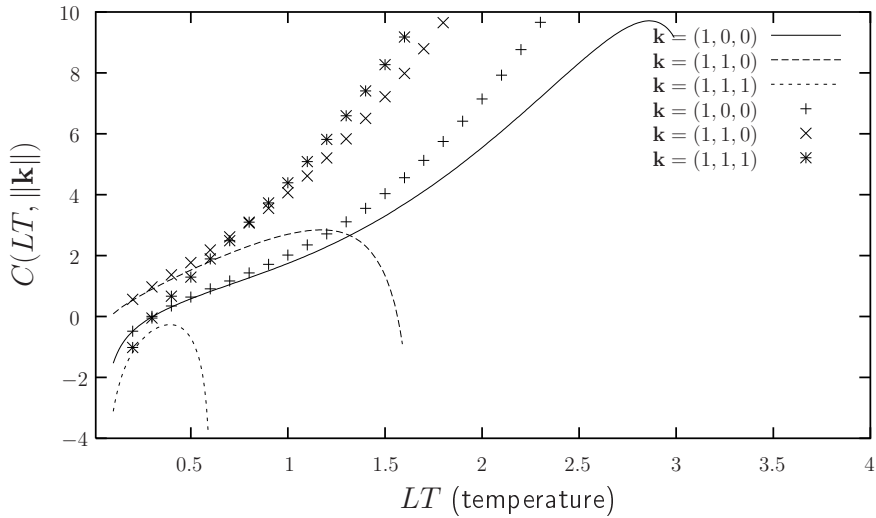}
        \caption{The quantity $C$ (\ref{eq:11_0004}) as a function 
	of $\len T$ for different twist
	vectors $\zt$. The lines show the function $C$ when all
	eigenvalues of the fluctuation spectrum are included in the
	corresponding proper time integral. The graphs plotted with the
	symbols ($+,\times,*$) show $C$ when the eigenvalues that become
	negative with rising temperature are neglected (see text).
	\label{abbtwiststemp}}
  \end{center}
  }
\end{figure}
\noindent
In figure \ref{abbtwiststemp} the quantity $C$ defined by 
eq.~(\ref{eq:11_0004}) which is part of the free energy (\ref{free-en}) 
is plotted as a function of $\len T$ for different twists. 
This quantity (and consequently also the free energy) goes to minus 
infinity when $\len T$ approaches the region where a single eigenvalue 
of the spectrum of the fluctuations becomes negative. This signals the
limit of validity of our calculation.
It is easy to see that the convergence condition (\ref{eq:05_0035}) can
not be valid for arbitrarily high temperatures: The twisted  
boundary conditions enforce a {\bf constant} flux 
$\|\zt\| = f \len \len_0$ in the 
plane of the twist, where flux is the product of area and (constant) 
field strength. Increasing the temperature $T$, i.e.~decreasing the torus
extension $L_0$, results in an increasing field strength until the
inequality  (\ref{eq:05_0035}) is violated.\\
\noindent
As can be seen from fig.~\ref{abbtwiststemp} the free energy of 
electric center vortices (temporal twist) increases with the 
temperature $T \len$. Consequently the electric center vortices 
become less and less important as the temperature increases. 
This result is consistent with the vanishing of the temporal string
tension above the deconfinement phase transition observed in lattice
calculations, given the fact that electric center vortices are
responsible for the temporal string tension.\\
\noindent
The points ($+,\times,*$) in figure \ref{abbtwiststemp} show $C$ as a
function of $T$ for the different twists when the negative eigenvalues
in the spectrum are neglected. 
The appearance of negative modes signals the infrared
instability of the perturbative vacuum. Neglecting the
negative eigenvalues the free energy is proportional to $T^4$ for 
large $T$, which corresponds to the \emph{Stefan-Boltzmann}-law 
of a free Bose gas.
\begin{figure}[!t]
  \vspace*{-.7cm}
  \textsf{
  \begin{center}
        \psfragscanon
        \psfrag{ 0}{$0$}
        \psfrag{ 0.2}{$0.2$}
        \psfrag{ 0.4}{$0.4$}
        \psfrag{ 0.6}{$0.6$}
        \psfrag{ 0.8}{$0.8$}
        \psfrag{ 1}{$1$}
        \psfrag{1.2}{}
        \psfrag{1.4}{$1.4$}
        \psfrag{1.6}{}
        \psfrag{1.8}{$1.8$}
        \psfrag{2}{}
        \psfrag{ 0.1}{$0.1$}
        \psfrag{ 0.12}{}
        \psfrag{ 0.14}{}
        \psfrag{ 0.16}{}
        \psfrag{ 0.18}{}
        \psfrag{ 0.2}{$0.2$}
        \psfrag{ 0.22}{}
        \psfrag{ 0.24}{}
        \psfrag{ 0.26}{}
        \psfrag{ 0.28}{}
        \psfrag{ 0.3}{$0.3$}
        \psfrag{L (Torusgroesse)}{\Large{$\len\laq$ (torus extension)}}
        \psfrag{exp(-F/T) morespace}{\Large{$\varexp{-F_{\zt}/T}$}}  
        \includegraphics[angle = 270,scale=.53]{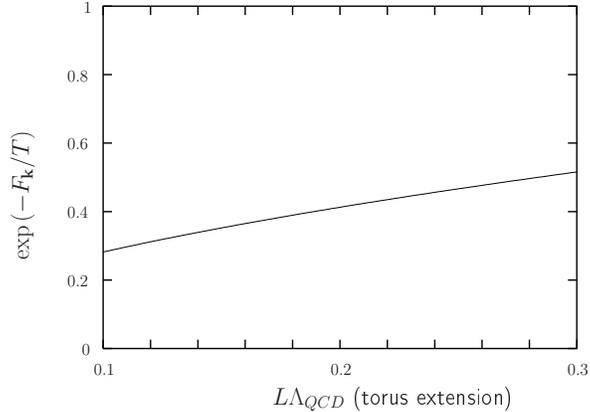}
        \caption{Creation probability of the thick electric vortex 
	as function of 	the torus extension. \label{abbtwiststemp2}}
  \end{center}
  }
\end{figure}
 
Let us consider the partition function 
$\Zs_{\zt}(\rt=0) = \varexp{-F_{\zt}/T}$ of the temporal 
twist $\zt = (1,0,0)$ on a torus with fixed ratio 
$\len / \len_0 = \len T$ as a function of the temperature $T$. 
From eq.~(\ref{eq:06_0002}) one obtains
\begin{align}
        \label{eq:06_0003}
        \Zs_{\zt} (T) &= 
	e^{- C (\len T)} \left( \len T \right)^{\frac{11}{6} \len T} 
	\left( \frac{T}{\laq} \right)^{-\frac{11}{6} \len T} \, , 
\end{align}
i.e.~for large temperature $T$ the partition function $\Zs_{\zt}$ 
decreases according to a power law with exponent $-\frac{11}{6} \len T$.
This is in qualitative agreement with the lattice results of 
ref.~\cite{deForcrand:2001nd}. There the partition function $\Zs_{\zt}$ 
has been calculated as a function of $T$ for different lattice 
geometries (corresponding to different values of $\len T$) and 
thereby the deconfinement phase transition was identified.
Below the critical temperature $\Zs_{\zt}$ is nearly one. Increasing 
the temperature $\Zs_{\zt}$ decreases and above the critical temperature 
drops to zero as predicted by eq.~(\ref{eq:06_0003}). The larger 
$\len T$ the sharper the drop. 
\newline
The partition function $\Zs_{\zt}$ in eq.~(\ref{eq:06_0003}) can also be
considered as a function of the torus extension $\len$ (with fixed 
value $\len T$). Figure \ref{abbtwiststemp2} shows $\Zs_{\zt}$ as a 
function\footnote{In the figure $\len$ is measured in units of 
$1/\laq$, i.e. with $\laq = 600$ MeV the value $\len \laq = 0.3$ 
corresponds to $\len \approx 0.1$ fm.} of $\len \laq$ for $\len T = 0.3$. 
This function can be 
interpreted as the creation probability of a thick vortex 
\cite{kovacs:00}. It monotonically increases with $\len \laq$. 
Although the present one-loop calculation is, strictly speaking,
reliable only for small $\len$ it certainly shows the right tendency of
$\Zs_{\zt}$ for increasing $\len$. Lattice calculations performed 
by Kov\'acs and Tomboulis \cite{kovacs:00} show that the center 
vortex creation probability $\Zs_{\zt} = \varexp{-F_{\zt}/T}$ approaches 
one for large $\len$, implying that the Yang-Mills vacuum can be 
considered as a condensate of thick center vortices.

With the above results at hand we can calculate the free energy 
of an electric flux $\zvec{e}$ and the expectation value of the 
\emph{Polyakov}-loop correlator, see eq.~(\ref{eq:02_0071}). 
Since the spectrum is invariant under spatial rotations contributions 
of the twists $\zt \in \{(1,0,0),(0,1,0),(0,0,1)\}$ are identical 
and we will write $\Zs_{\zt} (\rt =0 ,\len_\mu) = \Zs (1)$ if 
$\zt \in \{(1,0,0),(0,1,0),(0,0,1)\}$ and 
$\Zs_{\zt} (\rt =0 ,\len_\mu) = \Zs(2)$ for 
$\zt \in \{(1,1,0),(0,1,1),(1,0,1)$ and 
$\Zs_{\zt} (\rt =0 ,\len_\mu) = \Zs(3)$ for 
$\zt = (1,1,1)$. The partition function for the electric flux 
$\zvec{e} = (1,0,0)$ is given by
\begin{align}
        \label{eq:05_0037}
        \Zs ({\zvec{e}},\rt=0,\len_\mu) =  
	e^{-\beta F ({\zvec{e}},\rt = 0 ,L_\mu)} =
        \frac{1 + \Zs(1) - \Zs(2) - \Zs(3)}
        {1 + 3\Zs(1) + 3\Zs(2) + \Zs(3)} \, .
\end{align} 
\begin{figure}[!t]
  \textsf{
  \begin{center}
        \psfragscanon
        \psfrag{llaq = 1spa}{\Large{\hspace*{-2cm}$\len\laq = 0.07$}}
        \psfrag{llaq = 2spa}{\Large{\hspace*{-2cm}$\len\laq = 0.1$}}
        \psfrag{llaq = 3spa}{\Large{\hspace*{-2cm}$\len\laq = 0.15$}}
        \psfrag{genauso}{}
        \psfrag{(a)}{\;\;\;\;\;\;\;\;(a)}
        \psfrag{(b)}{(b)}
        \psfrag{0}{$0$}
        \psfrag{0.5}{}
        \psfrag{-0.5}{}
        \psfrag{1}{$1$}
        \psfrag{1.5}{}
        \psfrag{2}{$2$}
        \psfrag{2.5}{}
        \psfrag{3}{$3$}
        \psfrag{3.5}{}
        \psfrag{4}{$4$}
        \psfrag{4.5}{}
        \psfrag{5}{$5$}
        \psfrag{6}{$6$}
        \psfrag{7}{$7$}
        \psfrag{8}{$8$}
        \psfrag{LT (Temperatur)}{\Large{$\frac{T}{\laq}$ (temperature)}}
        \psfrag{Zk(1)}{\Large{$\Zs(1)$}}
        \psfrag{Ze(1)}{\Large{$\Zs(\zvec{e}=(1,0,0))$}}
        \includegraphics[angle = 270,scale=.53]{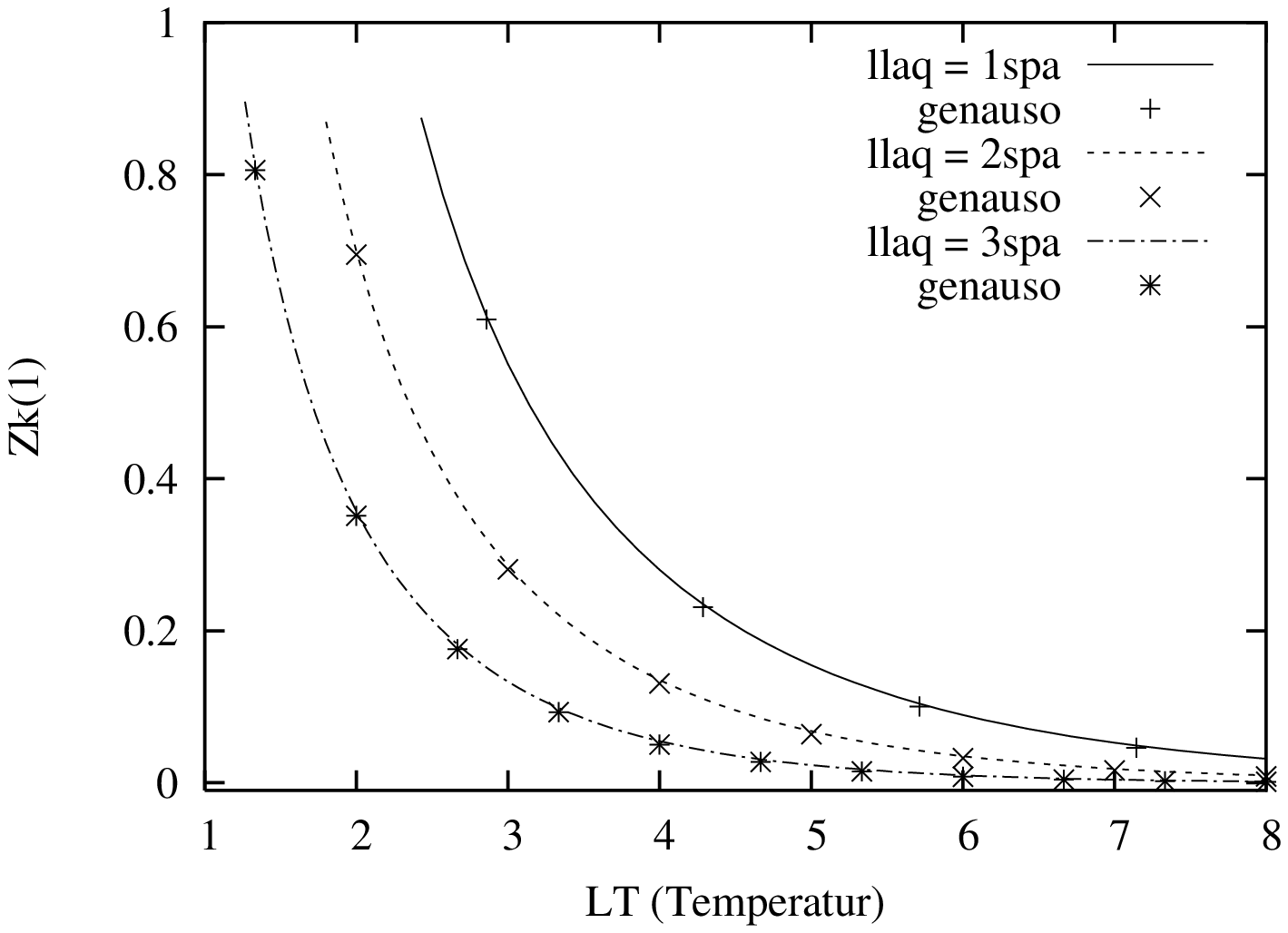}
        \includegraphics[angle = 270,scale=.53]{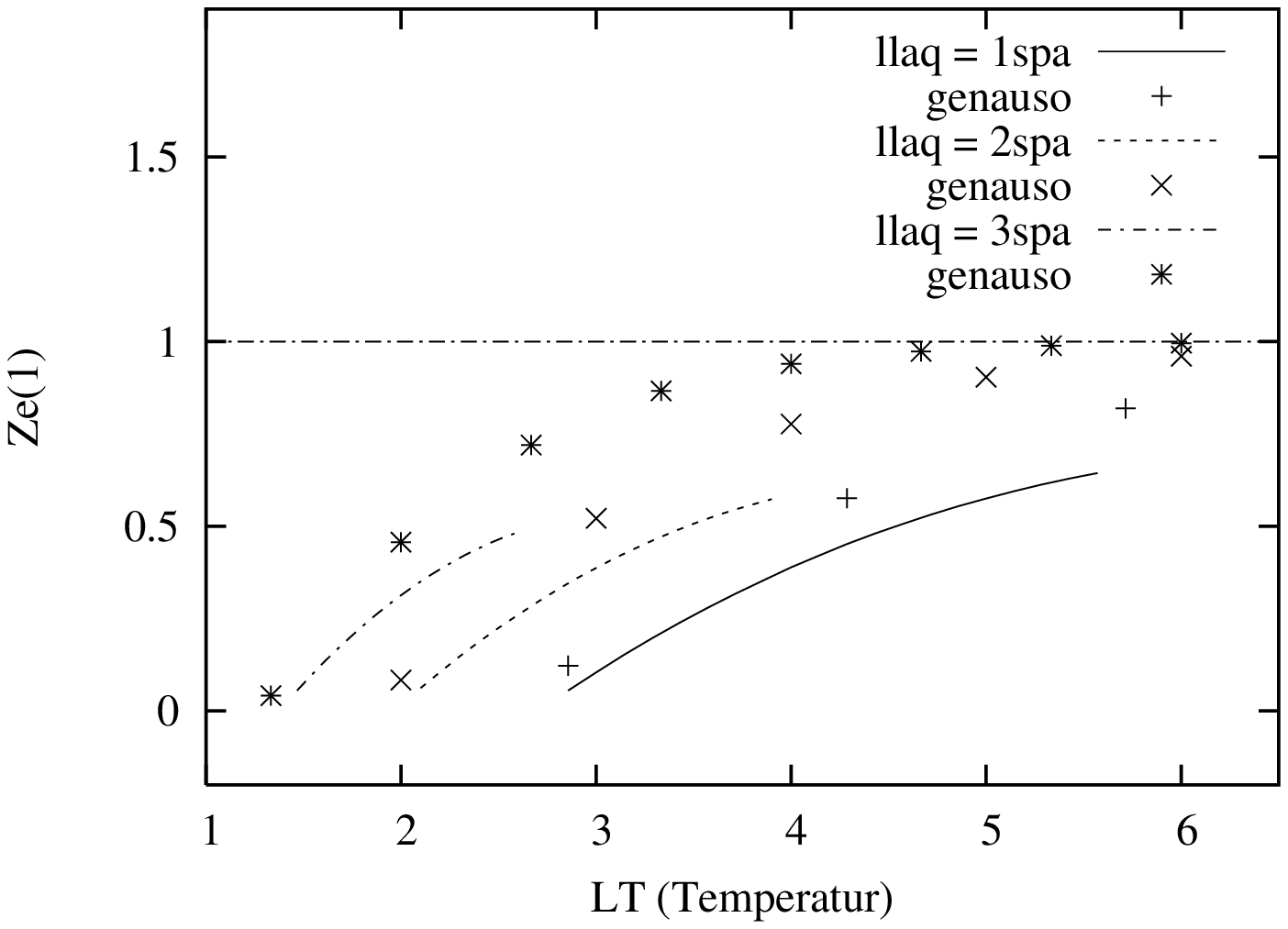}\medskip\\
        \includegraphics[angle = 0,scale=.9]{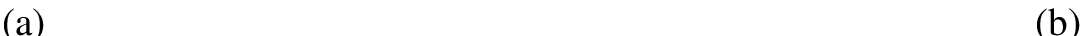}
       \caption{Partition function of (a) temporal twist $\zt=(1,0,0)$
       and (b) electric flux $\zvec{e} = (1,0,0)$ as function of
       temperature. The full lines give the results under
       consideration of all eigenvalues. The plots with $+,\times,*$
       show the results when the negative eigenvalues are neglected as
       in fig.~\ref{abbtwiststemp}.
       \label{abbtwiststemp3}} 
  \end{center}
  }
\end{figure}
\noindent
In figure \ref{abbtwiststemp3} the temperature dependences of 
$\Zs(1)$ (a) and $\Zs(\zvec{e}=(1,0,0))$ (b) are plotted for 
different values of $\len \laq$. Obviously, $\Zs(1)$ as a function of 
$T$ decreases with increasing temperature $T$ and for larger $\len$ it
becomes steeper at low temperatures $T$. In the range of temperature, 
where our calculation is valid, one observes a dual behaviour 
between $\Zs(\zvec{e})$ and $\Zs(1)$. This is expected since 
$\Zs(\zvec{e})$ emerges from $\Zs(i) \, , \, i = 1,2,3$ through a 
$Z(2)$ Fourier transform. Qualitatively, the same dual behaviour is 
observed in the lattice calculation \cite{deForcrand:2001nd}.
Unfortunately, the perturbative ansatz has the problem that 
the one-loop approximation is valid for high temperatures. 
On 
the other hand for the calculation of $\Zs(\zvec{e})$ all possible 
twists $\zt$ are needed - and
especially $\Zs(3)$ is known only for $T/\laq < 0.6/(L\laq)$. 
Therefore, $\Zs(\zvec{e})$ can hardly be considered in the scope of the
one-loop approximation.

\section{Conclusions}

In this paper we have carried out a one loop calculation of the free
energy of fat center vortices in $SU(2)$ Yang-Mills theory as function
of the temperature. The fat center vortices were induced by imposing
twisted boundary conditions to the gauge field defined on a four-torus.
These boundary conditions, in turn, were realized by Abelian background
fields which describe fat center vortices whose flux is homogeneously
distributed over the whole 4-torus. Accordingly the fluctuations around
the Abelian background fields have to satisfy quasi periodic boundary
conditions. For arbitrary combinations of electric and magnetic center 
flux, i.e.~for arbitrary twist, negative eigenmodes of the fluctuations  
appear. Such modes can be avoided for purely spatial or purely temporal
twists. In the case of purely spatial twist the free
energy is nearly independent of the temperature in agreement with
lattice results \cite{deForcrand:2001nd}. More interesting is the 
case of purely temporal twist.
Unfortunately in this case the range of validity of our calculations is
rather restricted: for low temperatures the perturbative ansatz 
is not valid while for high temperatures negative modes appear 
in the fluctuation spectrum. But nevertheless in the range of validity
of the present calculation our results are in 
qualitative agreement with the lattice results given in 
\cite{deForcrand:2001nd}. Furthermore, the creation probability 
of a thick vortex in dependence on the torus extension $L$ has been 
calculated and is in agreement with the lattice results given in 
\cite{kovacs:00}.

To summarize we have been able to (at least qualitatively) reproduce 
the lattice results obtained in refs.~\cite{deForcrand:2001nd,kovacs:00} by 
an one-loop calculation in continuum Yang-Mills theory.


\section{Acknowledgments}

The authors are grateful to L.~v.~Smekal and  K.~Langfeld 
for helpful discussions. 
This work has been supported by the Deutsche Forschungsgemeinschaft 
under grant DFG-Re 856/4-2.


\appendix

\section{Spectrum of the fluctuation operators $M_{n1}$ and $M_{n2}$}
\label{app_spec}

As pointed out in section \ref{spectrum} the operator $\M_n$, see 
eq.~(\ref{eq:03_0032}), can be written as the sum
\begin{align*}
   \M_n = 
   \left.\M_{n} \right|_{\img(\F)} + 
   \left.\M_{n} \right|_{\ker(\F)} \, .
\end{align*}
In the operator $\left.\M_{n} \right|_{\img(\F)}$ the 
constant term $\frac{2\pi\Q_\lambda}{\len_\lambda}$ can be absorbed 
in a constant shift $y_\mu$ of $x_\mu$:
\begin{align*}
\frac{2\pi\Q_\lambda}{\len_\lambda}|_{\img(\F)} = 
\pi\F_{\lambda\nu} y_\nu \, .
\end{align*}
\noindent 
The spectrum and the eigenfunctions of the operators
\begin{align*}
\left.\M_{n} \right|_{\img(\F)} = 
\left. \left(\rez{i}\dau\lambda - \pi\F_{\lambda\nu} ( x_\nu - y_\nu )
       \right)^2
       \right|_{\img(\F)} 
       \quad \mbox{and} \quad
       \left.(\left.\M_{n} \right|_{\img(\F)} \delta^{\mu\nu} - 
       4\pi i \F^{\mu\nu})\right|_{\img(\F)}
\end{align*}
\noindent 
have been calculated in \cite{vbaal:84}. 
In table \ref{tab01} we quote the result for the spectrum, 
\begin{table}[h]
\begin{align*}
       \begin{array}{l|l|l|l}
        \mathrm{operator }&\mathrm{eigenvalue}&
	\mathrm{degeneracy}&\mathrm{parameter}\\
	\hline
        \left.\M_{n} \right|_{\img(\F)} &
	2\pi f(2m + 1) & 2 \tilde e &  m = 0,1,2,\ldots \\
	\hline
        \left.(\left.\M_{n} \right|_{\img(\F)} \delta^{\mu\nu} - 
	 4\pi i \F^{\mu\nu})\right|_{\img(\F)} 
         & -2\pi f & 2\tilde e \\ 
         &  2\pi f & 6\tilde e \\ 
         & 2\pi f(2m + 1) & 8\tilde e &  m = 1,2,3,\ldots 
       \end{array}
\end{align*}
\caption{Spectrum of $\left.\M_{n} \right|_{\img(\F)}$ and 
         $\left.(\left.\M_{n} \right|_{\img(\F)} \delta^{\mu\nu} - 
	 4\pi i \F^{\mu\nu})\right|_{\img(\F)}$. \label{tab01}}
\end{table}
where we have introduced the number 
\begin{align}
\label{eq:001}
f = \frac{\|\rt\|}{L^2} + \frac{\|\zt\|}{L L_0}
\end{align} 
and the integer 
$\tilde e$ which is the greatest common divisor of the integers 
$\rt_1,\rt_2,\rt_3,\zt_1,\zt_2,\zt_3$:
\begin{align}
\label{eq:002}
\tilde e = \gcd{(\rt_1,\rt_2,\rt_3,\zt_1,\zt_2,\zt_3)} \, .
\end{align} 
Note we have chosen $\len_1=\len_2=\len_3=\len$ and restricted ourselves
to either $\rt = 0$ or $\zt=0$. 


The eigenfunctions of $\left.\M_n \right|_{\ker(\F)}$ 
are given by plane 
waves $\exp(-i p_\mu x_\mu)$. To get its eigenvalues 
one has to analyze the periodicity properties of the functions
$c_\mu(x)$ on the space $\ker(\F)$, or more precisely on 
$\left. \Lat \right|_{\ker(\F)} = \Lat \cap \ker(\F)$ which is a 
planar Bravais lattice. Let the lattice $\left. \Lat \right|_{\ker(\F)}$ be
generated by the two vectors $\{\zvec{\ab}_1, \zvec{\ab}_2\}$ (which 
cannot be chosen orthogonally in general). From eq.~(\ref{eq:03_0026}) 
one obtains
\begin{align}
        \label{eq:03_0026A}
       \ce_\lambda\left(x + l_1 \zvec{\ab}_1 + l_2 \zvec{\ab}_2 \right) =
        \varexp{i\pi
        \sum_{j=1}^2 l_j^2 n_j}
	\ce_\lambda(x) \, ,
\end{align}
\noindent
where 
$n_j=\sum_{\mu<\nu} (\zvec{\ab}_j)_\mu n_{\mu\nu} (\zvec{\ab}_j)_\nu$. 
If $n_j$ is an odd number the function $c_\lambda$ is antiperiodic 
under a shift by $\zvec{\ab}_j$, otherwise it is periodic. The 
possible momentum vectors can be written down in terms of the 
basis vectors $\{\zvec{\bb}_1,\zvec{\bb}_2\}$ of the inverse 
lattice: 
\begin{align}
  \label{eq:03_0054}
        \zvec{p}(k,l) = (k+\Delta_1) \zvec{\bb}_1 +
        (l +\Delta_2) \zvec{\bb}_2, \ha k,l  \in \mathbbm{Z} \, ,
\end{align}
\noindent 
where $\Delta_j=0$ if $n_j$ is even and $\Delta_j=1/2$ if $n_j$ is odd
and the basis vectors 
$\zvec{\bb}_j$ of the inverse lattice are defined by the relation
\begin{align}
\zvec{\ab}_i\cdot\zvec{\bb}_j = 2\pi\delta_{ij} \, .
\end{align}
For given twist vector $\rt$ or $\zt$, resp., it is straightforward to
calculate the vectors $\zvec{\ab}_j$ and $\zvec{\bb}_j$. The results are
summarized in table \ref{tab_05}.
\begin{table}[ht]
    \begin{center}
        $\zt=0 \,$: \\ \vspace{0.3cm}
	\begin{tabular}{c|c|c|c|c|c|c}
          $\rt$ & 
	  $\zvec{\ab}_1$ &
	  $\zvec{\ab}_2$ &
	  $\zvec{\bb}_1$ &
	  $\zvec{\bb}_2$ &
	  $\Delta_1$ &
	  $\Delta_2$ 
	  \\ \hline
	  $(1,0,0)$ & 
	  $(L_0,0,0,0)$ &
	  $(0,L,0,0)$ &
	  $\frac{2 \pi}{L_0}(1,0,0,0)$ &
	  $\frac{2 \pi}{L}(0,1,0,0)$ &
	  $0$ &
	  $0$ 
	  \\
	  $(1,1,0)$ &
	  $(L_0,0,0,0)$ &
	  $(0,L,L,0)$ &
	  $\frac{2 \pi}{L_0}(1,0,0,0)$ &
	  $\frac{2 \pi}{2 L}(0,1,1,0)$ &
	  $0$ &
	  $0$ 
	  \\
	  $(1,1,1)$ &
	  $(L_0,0,0,0)$ &
	  $(0,L,L,L)$ &
	  $\frac{2 \pi}{L_0}(1,0,0,0)$ &
	  $\frac{2 \pi}{3 L}(0,1,1,1)$ &
	  $0$ &
	  $1/2$
        \end{tabular} 
	\\ \vspace{0.3cm}
	$\rt = 0 \,$: \\ \vspace{0.3cm}
	\begin{tabular}{c|c|c|c|c|c|c}
          $\zt$ & 
	  $\zvec{\ab}_1$ &
	  $\zvec{\ab}_2$ &
	  $\zvec{\bb}_1$ &
	  $\zvec{\bb}_2$ &
	  $\Delta_1$ &
	  $\Delta_2$ 
	  \\ \hline
	  $(1,0,0)$ & 
	  $(0,0,L,0)$ & 
	  $(0,0,0,L)$ & 
	  $\frac{2 \pi}{L}(0,0,1,0)$ &
	  $\frac{2 \pi}{L}(0,0,0,1)$ &
	  $0$ &
	  $0$ 
	  \\
	  $(1,1,0)$ &
	  $(0,0,0,L)$ &
	  $(0,L,-L,0)$ &
	  $\frac{2 \pi}{L}(0,0,0,1)$ & 
	  $\frac{\pi}{L}(0,1,-1,0)$ & 
	  $0$ &
	  $0$ 
	  \\
	  $(1,1,1)$ &
	  $(0,-L,L,0)$ &
	  $(0,0,-L,L)$ &
	  $\frac{2 \pi}{3 L}(0,-2,1,1)$ &
	  $\frac{2 \pi}{3 L}(0,-1,-1,2)$ &
	  $0$ &
	  $0$
        \end{tabular} 
   \caption{\label{tab_05} Generating vectors of 
   $\left. \Lat \right|_{\ker(\F)}$ and its dual lattice.}
   \end{center}
\end{table}
The eigenvalues of the operator $\left.\M_{n} \right|_{\ker(\F)}$ 
in terms of $\zvec{\bb}_j$, 
$\Delta_j$ and $\Q_\lambda$ read:
\begin{align*}
        \lambda_{(k,l)} = \left((k+\Delta_1) \zvec{\bb}_1 +
        (l +\Delta_2) \zvec{\bb}_2 + \left . 
	\frac{2\pi \Q_\lambda}{\len_\lambda}\right|_{\ker(\F)}
	\right)^2 \, , \ha k,l  \in \mathbbm{Z} \, .
\end{align*}


\section{Free energy}
\label{proper_time}

To evaluate the proper time integral (\ref{eq:05_0002}) 
we start with the simplest case of zero twist $n_{\mu\nu}= 0$. The
corresponding eigenvalues are given in 
eqs.~(\ref{eq:03_0059},\ref{eq:03_0060}).
Corresponding to the different types of eigenvalues we will split the
proper time integral $\log\det\M_A^{\mu \nu}(n_{\mu\nu} =  0)=I_1+I_2$, 
where $I_1$ and $I_2$ contain all eigenvalues from equations 
(\ref{eq:03_0059}) and (\ref{eq:03_0060}), respectively:
\begin{align}
  \label{eq:05_0006}
        I_1 &= -4 \int^\infty_{\rez{\Lambda^2}}
        \frac{\de \tau}{\tau}\sum_{l \in {\mathbbm
        Z}^4\backslash \{0\}}\varexp{-\tau\sum_{\mu=0}^3\left(\frac{2\pi
        l_\mu}{\len_\mu}\right)^2} \, , \\
  \label{eq:05_0007}
        I_2 &= -8 \int^\infty_{\rez{\Lambda^2}}
        \frac{\de \tau}{\tau}\sum_{l \in {\mathbbm
        Z}^4}\varexp{-\tau\sum_{\mu=0}^3\left(\frac{2\pi
        (l_\mu - \Q_\mu)}{\len_\mu}\right)^2} \, . 
\end{align}
\noindent 
Under a suitable choice of $\Q_\mu$ all eigenvalues in $I_2$ are
positive. 
Using the definition of the theta function
\begin{align}
        \label{eq:05_0008}
        \theta(z, t) = \sum_{n\in {\mathbbm Z}}\varexp{\pi i n^2 t +
        2\pi i n z}, \ha \Im t > 0, z \in {\mathbbm C},
\end{align}
\noindent 
and substituting $\tau$ by $\frac{4 \pi}{\len^2} \tau$ one obtains
\begin{align}
        \label{eq:05_0009}
        -\log\det M_A(n=0) = -I_1 - I_2 = 
        \int^\infty_{\frac{4 \pi}{\len^2 \Lambda^2}}
        \frac{\de \tau}{\tau}\Bigg[4\Bigg[\prod_\mu \theta\left(0,
        i\frac{\tau}{\pi} \left(\frac{2\pi}{\len_\mu}\right)^2\right) -1\Bigg]
        +8 \so(\tau) \Bigg] \, , 
\end{align}
where we have introduced the function
\begin{align}
	\label{eq:05_0009A}
	\so (\tau)&:= 
	\prod_\mu\theta\left(
	-i \tau \left(\frac{\len}{\len_\mu}\right)^2 \Q_\mu, 
	i\tau \left(\frac{\len}{\len_\mu}\right)^2 \right)
	e^{-\tau \pi \left(\frac{\len}{\len_\mu}\right)^2 \Q_\mu^2} 
	\, .
\end{align}
For non-zero twist ($n_{\mu \nu} \neq 0$) the calculations are done 
in an analogous fashion. One finds 
\begin{align}
        \label{eq:05_0011a}
	\log\det\M_A^{\mu \nu}(n_{\mu\nu} \neq  0)=I_1+I_3 \, ,
\end{align}
\noindent
where $I_1$ is defined by eq.~(\ref{eq:05_0006}) and contains all the
eigenvalues given in the lowest line of table (\ref{tab02}), and 
\begin{align}
        \label{eq:05_0012}
        I_3 &=  - 2 \int^\infty_{\frac{4\pi}{\len^2\Lambda^2}} 
        \frac{\de \tau}{\tau}
	\left[\tilde e\frac{e^{-2 f \len^2 \tau} 
        + 2e^{- f \len^2 \tau} + 1}{(1-e^{- f \len^2 \tau})}
        e^{\rez 2 f \len^2 \tau}
        \sum_{k,l\in\mathbbm{Z}}e^{-\frac{\tau
         \len^2}{4\pi}\left(\zvec{p}(k,l) - \zvec{\q}\right)^2}
	\right] \, 
\end{align}
\noindent
and 
\begin{align}
        \label{eq:05_0012A}
        \log \frac{\det M_A(n)}{\det M_A(n=0)} = I_3 + 
	8 \int^\infty_{\frac{4\pi}{\len^2\Lambda^2}} 
        \frac{\de \tau}{\tau} \so(\tau) \, .
\end{align}
\noindent 
In a similarly way one obtains for the determinant of $\M_{gh}$:
\begin{align}
        \label{eq:05_0012B}
        \log\frac{\det \M_{gh}(n)}{\det \M_{gh}(n=0)} =
         - 2 \int^\infty_{\frac{4\pi}{\len^2\Lambda^2}} 
        \frac{\de \tau}{\tau}\left[\tilde e
	\frac{e^{-\rez 2 f \len^2 \tau}}{(1-e^{- f \len^2 \tau})}
        \sum_{k,l\in\mathbbm{Z}}e^{-\tau\frac{L^2}{4\pi}\left(\zvec{p}(k,l) -
         \zvec{\q}\right)^2} - \so(\tau)\right] \, .
\end{align}
\noindent 
The free energy 
\begin{align}
        \label{eq:05_0020}
	F_{\zt}(\rt) = -T \log \Z_{\zt}(\rt)
\end{align}
\noindent
then becomes
\begin{align}
        \label{eq:05_0020A}
	\frac{F_{\zt}(\rt)}{T} &= 
	-\left( -S_0-\rez{2}\log\det \tilde \M_A^{\mu\nu}
        + \log\det\tilde \M_{gh} \right) \\
        \label{eq:05_0020A1}
	&=
	\frac{2\pi^2}{g^2} \frac{\left(f \len^2\right)^2}{\len T} - 
	\int^\infty_{\frac{4\pi}{\len^2\Lambda^2}} 
        \frac{\de \tau}{\tau} I(\tau) \, , 
\end{align}
where we have introduced the abbreviation
\begin{align}
        \label{eq:05_0020B}	
	I(\tau) & = 
	\left[\tilde e \frac{e^{-2 f \len^2 \tau} + 1
	}{(1-e^{- f \len^2 \tau})} e^{\rez 2 f \len^2 \tau}
         \sum_{k,l\in\mathbbm{Z}}e^{-\tau\frac{L^2}{4\pi}\left(\zvec{p}(k,l) -
         \zvec{\q}\right)^2} - 2\so(\tau)\right] \, . 
\end{align}
\noindent
The function $I(\tau)$ is regular at $\tau=0$ 
$$
I(\tau=0) = \frac{11}{12} \frac{\left( f \len^2\right)^2}{\len T}
$$
so that the ultraviolet singularity in the proper time integral can be
easily identified by writing
\begin{align}
        \label{eq:05_0020C}
	\int^\infty_{\frac{4\pi}{\len^2\Lambda^2}} 
        \frac{\de \tau}{\tau} I(\tau) &=
	\int^\infty_{\tau_{min}} 
        \frac{\de \tau}{\tau} I(\tau)
	+ I(0) \left( \log\frac{\tau_{min}}{4\pi} 
	           + \log{\len^2\Lambda^2} \right) 
	+ \int^{\tau_{min}}_{\frac{4\pi}{\len^2\Lambda^2}} 
	\frac{\de \tau}{\tau} (I(\tau) - I(0) ) \, ,
\end{align}
where we introduced some small but finite $\tau_{min}>0$. The integrals
on the r.h.s.~are finite and $\log{\len^2\Lambda^2}$ is the familiar
logarithmic singularity. As usual this singularity is absorbed into the
renormalized coupling constant defined at some energy scale $\mu$
\begin{align}
        \label{eq:05_0021}
        \rez{g_R^2(\mu^2)} := \rez{g^2} +
        \frac{11}{24\pi^2} \log\frac{\mu^2}{\Lambda^2} \, .    
\end{align}
\noindent 
In the present context it is convenient to identify this scale with the
temperature $\mu = T$. Inserting eq.~(\ref{eq:05_0020C}) into 
eq.~(\ref{eq:05_0020A}) the free energy in terms of the renormalized
coupling constant (\ref{eq:05_0021}) becomes
\begin{align}
        \label{eq:05_0030}
        \frac{F_{\zt}(\rt , \len T)}{T} &= 
	\frac{2\pi^2}{g_R^2(T^2)} 
	\frac{\left(f \len^2\right)^2}{ \len T}
        + \frac{11}{12} \frac{\left(f \len^2\right)^2}{\len T} 
	\log\frac{1}{\len^2 
        T^2} + C (f \len^2, \Q_\mu, \len T) \, , 
\end{align}
\noindent
where we have introduced the function
\begin{align}
        \label{eq:11_0004}
	C & =-\frac{11}{12} \frac{(f \len^2)^2}{\len T}
        \log\frac{\tau_{min}}{4\pi}
        -\int^\infty_{\tau_{min}} 
        \frac{\de \tau}{\tau} I(\tau) - 
        \int^{\tau_{min}}_0 \frac{\de \tau}{\tau}
	\left[ I(\tau) - I(0) \right] 
\end{align}
\noindent
which is independent of $\tau_{min}$ and can be determined 
numerically with very high precision. The function $I(\tau)$ is defined
in eq.~(\ref{eq:05_0020B}).

Introducing the parameter $\laq$ by
\begin{align}
        \label{eq:06_0001}
        \rez{g_R^2(T^2)} := \frac{11}{12 \pi^2} \log \frac{T}{\laq}    
\end{align}
we obtain for the free energy in terms of $\laq$:
\begin{align}
        \label{eq:06_0002}
        \frac{F_{\zt}(\rt , \len T)}{T} &= 
	- \frac{11}{6} \frac{\left(f \len^2\right)^2}{\len T} 
	\log \left( \len \laq \right)
        + C (f \len^2, \Q_\mu, \len T) \, .
\end{align}




\end{document}